# Impact Of Emotions on Information Seeking And Sharing Behaviors During Pandemic

Smitha Muthya Sudheendra
University of Minnesota,Twin Cities
muthy009@umn.edu

Jisu Huh
University of Minnesota,Twin Cities
jhuh@umn.edu

Hao Xu
The University of Melbourne, Melbourne
hao.xu@unimelb.edu.au

Jaideep Srivastava
University of Minnesota,Twin Cities
srivasta@umn.edu

## Abstract

We propose a novel approach to assess the public's coping behavior during the COVID-19 outbreak by examining the emotions. Specifically, we explore: (1) changes in the public's emotions with the COVID19 crisis progression; and (2) impacts of the public's emotions on their information-seeking, information-sharing behaviors, and compliance with stay-at-home policies. We base the study on the appraisal tendency framework, detect the public's emotions by fine-tuning a pre-trained RoBERTa model, and cross-analyze third-party behavioral data. We demonstrate the feasibility and reliability of our proposed approach in providing a large-scale examination of the public's emotions and coping behaviors in a real-world crisis: COVID-19. The approach complements prior crisis communication research, which is mainly based on self-reported, small-scale experiment and survey data. Our results show that anger and fear are more prominent than other emotions experienced by the public at the pandemic's outbreak stage. Results also show that the extent of low-certainty and passive emotions (e.g., sadness, fear) was related to increased information seeking and information-sharing behaviors. Additionally, degrees of both high-certainty (e.g., anger) and low-certainty (e.g., sadness, fear) emotions during the outbreak were found correlated to the public's compliance with stay-at-home orders.

## Introduction

COVID-19 was declared a Public Health Emergency of International concern in January 2020 by the World Health Organization (WHO). In the U.S., the first case was reported on January 20, 2020, the first known deaths occurred in February, the federal government declared a national emergency in March, and by mid-April, cases had been confirmed in all 50 states.

Previous studies show that public health crises, especially infectious disease outbreaks, can not only pose severe threats to the health of the population but also endanger social well-being (Morens and Fauci 2013). In a state of distress, it was important to understand the necessary precautionary steps to be taken based on the behaviors of the public and the nature of spread of the virus.

During crises, emotions function as "one of the anchors of the publics' interpretation of the unfolding and evolving events" (Jin, Pang, and Cameron 2012 p. 268). According to psychological research on stresses of harm or loss, there is a common tendency for people in crises to act on their emotions and take emotion-corresponding actions to cope with stressful events (Lazarus 1991). In major public health crises caused by infectious diseases, for example, emotions and emotion-based information, fueled by the infectious nature of the diseases, have the potential to spread across a broad population quickly (Dredze, Broniatowski, and Hilyard 2016), and could influence people's various coping behaviors (Jin et al. 2020). Given the critical importance of people's communicative behaviors, sharing information and disease-prevention behaviors in saving lives during public health crises, it is imperative to build a solid understanding of how the general public emotionally reacts to a public health crisis and how their emotions impact crisis coping behaviors.

Previously, computer scientists have studied social media to assess emotions during health crises and social scientists have studied the phenomenon using surveys and experiment data. In our study, we want to bridge the gap between the two fields by using a computational approach grounded in social science theory to examine the public's emotional reactions to the COVID-19 outbreak crisis in the U.S., and the impact of such emotions on their communicative behaviors, including information seeking and sharing, and compliance to stay-at-home order.

**Definitions:**

In this paper, we use a few terms particular to strategic health crisis communication. The terms are discussed below:

- Appraisal Tendency Framework: The Appraisal Tendency Framework (ATF) contains two processes (Lerner and Keltner 2001). First, discrete emotions arise from how different situations are appraised by individuals, which is driven by different "appraisal themes" underlying specific events (Lazarus 1991). Second, the emotions evoked would be carried over to subsequent events and determine the cognitive patterns of how emotions would influence sequential cognitive and conative coping (Lerner and Keltner 2001).
- High vs. low-certainty emotions: Certainty arises when individuals have reliable knowledge that the situation could, to a large extent, be predictable as it evolves (Smith and Ellsworth 1985). High-certainty emotions (e.g., anger, happiness) are likely to be evoked when individuals feel certain about the situation. In contrast, low-certainty emotions (e.g., fear, sadness) can be evoked by a sense of uncertainty about the situation.
- Agnostic vs. passive emotions: In crisis situations, controllability and responsibility appraisals reflect the degree to which individuals blame others for causing or mishandling the crisis (Coombs 1998). Agonistic emotions (e.g., anger) are likely to be evoked when people appraise the involved organization as having control over the crisis or attribute blame to the organization for causing or mishandling the crisis. In contrast, appraisals of low responsibility or low control by the involved organizations are more related to passive emotions (e.g., sadness, fear) (Smith and Ellsworth 1985).
- Coping behaviors: Coping behaviors describe actions people take to manage crisis situations. In this study, we focus on the public's communicative coping behaviors, including information seeking and sharing, and their compliance with governments' stay-at-home orders.

## Background

### Emotion Detection during Crisis

Emotion detection using social media during crisis situations has been of keen interest to researchers. Emotion detection using various classifiers have been implemented to analyze people's emotional responses during various crisis situations like natural disasters (e.g., earthquakes, hurricanes and tsunamis), public health crises (e.g., COVID-19) and human rights crises.

For example, in existing studies, emotion detection using multi-naive Bayes (MNB) with n-grams was implemented to understand the major emotions during earthquakes (Vo and Collier 2013). The authors considered six emotions, including calmness, unpleasantness, sadness, anxiety, fear, and relief, identified by Tokuhisa, Inui, and Matsumoto (2008). The results showed that fear and anxiety were prevalent after the earthquakes. Also, emotion detection using the RNN model was implemented to understand emotions of different groups of Twitter users and their tweeting patterns during Hurricane Dorian (Li and Fox 2020). Their study adopted six primary emotions identified by Ekman (1970), including anger, disgust, fear, happiness, sadness and surprise.

Since the inception of COVID-19, some studies have been conducted to understand the public's emotions during the pandemic. For example, Transfer Learning and Robustly Optimized BERT Pretraining Approach (RoBERTa), with an accuracy of 80.33%, was implemented to understand the emotions contained in tweets from all over the world (Choudrie et al. 2021). Also, using an analytical tool, CrystalFeel, researchers studied tweets from late January to early April to understand the emotion trend during the COVID-19 outbreak (Lwin et al. 2020). Their study considered four emotions, including fear, anger, sadness, and joy, based on Plutchik's Wheel of Emotions (Plutchik 1991) and showed that major emotions expressed changed from fear to anger.

Our study aims to understand not only the most prevalent emotions during the COVID-19 outbreak, but also the potential impacts of these emotions on the public's coping behaviors. Informed by the ATF, a social psychology theory, we consider emotions as mental states emerging from the appraisal of one's situation. Accordingly, various emotions have differences and similarities in cognitive appraisal dimensions, which could systematically influence one's subsequent behaviors (Lazarus 1991; Smith and Ellsworth 1985). Considering the context of COVID-19 outbreak, we detect three negative emotions, including anger, sadness, and fear, as well as positive emotions. We fine-tuned RoBERTa model for emotion detection.

### Impacts of Emotions on the Public's Coping Behaviors

During a disease outbreak, people would engage in various behaviors that could help them better cope with the situation. Prior research has shown that people seek and share crisis-related information during a public health crisis, in order to have a better understanding of the potential risks and threats (Lee and Jin 2019). In such volatile situations, people also make decisions on whether or not to follow protective actions ordered by governments and health organizations (Jin et al. 2020). Drawing upon the ATF, we investigate the impact of people's emotions on their information-seeking and information-sharing behaviors, and their compliance with the stay-at-home orders during the COVID-19 outbreak.

### Impacts on Information-Seeking Behavior

Different emotions are related to different levels of certainty appraisals. During a public health crisis, people's perceived uncertainty can increase when they face existential threats or think existing information to be inadequate or even mistaken to help them cope with the crisis (Yang and Chu 2018). For example, anger is a high-certainty emotion, as it can be evoked when individuals appraise the ongoing situation as certain, whereas fear and sadness are low-certainty emotions often evoked by uncertainty about the situation.

According to the ATF, in a public health crisis, people with low-certainty emotions are likely to have greater motivations to make sense of the ongoing situation and thus actively seek for relevant information in order to reduce uncertainty (Jin, Fraustino, and Liu 2016). Meanwhile, people with high-certainty emotions tend to be confident about ongoing situations (Tiedens and Linton 2001), and thus they are less likely to engage in information seeking.

### Impacts on Information-Sharing Behavior

Emotions in crises are also related to controllability and responsibility appraisals, which refer to the tendency of blaming others for causing and mishandling the crisis. (Coombs 1998). Appraisals of high levels of human control and others' responsibility can lead to agonistic emotions, such as anger toward the wrongdoing individual or organization (Smith and Ellsworth 1985). In contrast, appraisals of low responsibility or low human control are related to passive emotions, like sadness or fear (Smith and Ellsworth 1985).

Considering the ATF, people with agonistic emotions are more likely to initiate proactive and aggressive behaviors because they believe that they can influence the situation (Lerner and Tiedens 2006). In the context of crises, it has been found that people with agnostic emotions, such as anger, tend to engage in crisis information-sharing behaviors, as compared to people with passive emotions (Jin, Fraustino, and Liu 2016).

Impacts on Social Distancing Compliance Behavior

It is important to understand the instinct of individuals to protect themselves during a crisis because taking recommended protective actions, such as sheltering in place or wearing appropriate personal protective equipment, can save lives. Studies have shown that compliance to recommended protective actions is more likely to occur when people perceive a high level of threats (e.g., Neuwirth, Dunwoody, and Griffin 2000; Peters, Ruiter, and Kok 2014).

According to the ATF, severe risks associated with crises are likely to evoke low-certainty emotions, such as anxiety and fear, which in turn can facilitate people's protective actions, because such actions can help people gain certainty (Freberg 2012). This is supported by a prior survey, which found that when individuals experienced fear and anxiety during a hypothetical terrorist attack, they were more likely to take protective actions (Jin, Fraustino, and Liu 2016).

# Method

In this section, we discuss the data, variables considered and analyze the coping behavior of the social media users during the pandemic.

### Data

The timeline considered for this paper is from March 1st, 2020 to April 25th, 2020. The data considered are of two types: social media data and third-party behavioral data. The social media considered is twitter. We considered the tweets collected by QCRI (Qazi, Imran, and Ofli 2020) using the AIDR tool. This dataset consisted of 51 million geo-tagged tweets, but due to Twitter API limitations, we randomly selected 1.75% to hydrate the tweets for our study for the timeline considered. Then, we filtered out the replies, deleted tweets, and tweets/retweets from removed accounts. The final tweet dataset consisted of 153,364 source tweets and 482,665 retweets.

The two third-party behavioral datasets are: Google Trends data and Google mobility dataset. The Google Trends data provide aggregated, normalized indexes that show the relative popularity of specific search queries in Google Search across various regions. The Google Trends data, for in this study, was retrieved as weekly indexes of English searches for the top three keywords: "coronavirus", "COVID", and "COVID-19" at the state level.

The COVID19 community mobility report by Google consisted of movement trends of the public over time broken down by the location across different categories of places like retail and recreation, groceries and pharmacies, parks, transit stations, workplaces, and residential. Google compared these numbers with the baseline days before the COVID-19 outbreak (January 3 to February 6, 2020) and formed a series of indexes indicating relative changes in community mobility. We retrieved the weekly mobility indexes at the state level.

### Variable Selection

#### The Public's Emotions

To understand the public's emotions, we analyzed the social media data (i.e., tweets from the public) during COVID-19. We used fine-tuned Robustly Optimized BERT Pretraining Approach (RoBERTa) (Liu et al. 2019) to detect the emotions in the tweets. The learning rate was set to 2e-5 over three epochs. The training dataset considered for this study was an augmented dataset of an already available combined set and additional keywords related to each emotion. The combined dataset of ISEAR, emotion-stimulus (Ghazi, Inkpen, and Szpakowicz 2015), and DailyDialog (Li et al. 2017) with sentences, each annotated with one discrete emotion label for the selected five emotions. This dataset was augmented with a list of emotion-related keywords from relatedwords.org, we consolidated a list containing the keywords related to each emotion, then excluded jargon and field-specific terminologies related to the emotions from the list. This augmentation created a new training dataset with more balanced classes. This model can analyze social media content that contains social media text and classify each tweet into one of the five emotion categories: anger, sadness, fear, positive emotion (e.g., gratitude, pride, or relief), and other emotions or no emotion. We consider positive emotions as a single class because we are more interested in the fine-grained negative emotions expressed during a public health crisis. This model achieved an F1 score of 0.83. Following the emotion detection, the relative prevalence of emotions for any given week at the state level was calculated as the percentage of the number of tweets classified into each emotion category out of the total number of tweets posted weekly in each state (ranging from 0 to 100%). This list of emotions by the state for the eight weeks acts as our independent variable.

**Information-Seeking Behavior**

During the COVID-19 pandemic, internet search was one of the primary methods of seeking relevant information (Lu and Reis 2021). Hence, we considered the state-level Google search trends data as a proxy measure to understand the information-seeking behavior, since Google Search accounted for approximately 90% of all U.S. web searches (Statcounter n.d.). The retrieved data was for three highest search terms: "coronavirus", "COVID", and "COVID-19". The indexes were normalized on a scale of 0-100 for each term and were calculated based on the location and given timeline. The data was then aggregated over all the terms to provide a list of values representing the relative intensity of COVID-related information seeking by region for the eight studied weeks.

**Compliance with Stay-at-Home Orders**

In the initial stages of the pandemic outbreak, the federal, state, and local governments in the U.S. implemented strict measures like stay-at-home orders, in order to decrease the chance of spreading the disease. The community-level people's movements can be inferred to understand the general public's compliance with the orders. Hence, we used two sets of Google Community Mobility indexes: (1) the index of movements in the residential areas and (2) the index of movements in the public areas, calculated as the average of the movement indexes of retail and recreation, groceries and pharmacies, transit stations, and workplaces.

**Information-Sharing Behavior**

Since its inception, Twitter has been a popular social media platform where people often express their opinions and thoughts. It is one of the most popular social media platforms in the U.S. and plays a vital role in American people's daily information exchange. Thus, Twitter data has been frequently used to study various social phenomena and human behaviors (Havey 2020; Steinert-Threlkeld 2018). During crises, posting tweets and retweeting others' posts about the situation are considered common information-sharing behaviors (Lee and Jin 2019). We developed a proxy measure for the public's information-sharing behaviors by calculating the weekly sums of the tweeting volume and retweeting volume by state and normalized them to represent relative changes over time in each state. The three lists, information sharing, information seeking, and compliance to follow stay-at-home order are the dependency variables.

**Political Leaning**

COVID-19 measurement is also a politicized issue in the U.S., as found in several national polls (Newport 2020; Pew Research Center 2020). People with different political ideologies have reacted differently to the information regarding COVID-19 and the measures taken (Gollwitzer et al. 2020). Hence, we use political leaning as a control variable in testing our hypothesis. We considered the 2020 presidential election results by state (FEC n.d.) as a proxy measure for each state's political leaning. The value of political leaning by state was calculated as the percentages of votes Biden/Harris received out of the total number of votes in each state (ranging from 0 to 100%).

**Understanding the Public's Coping Behaviors**

Based on the Appraisal Tendency Framework and existing research on crisis emotions (e.g., Jin, Fraustino, and Liu 2016; Yang and Chu 2018), we proposed a series of hypotheses on the relationships between the public's emotions and coping behaviors, including their information-seeking, information-sharing, and compliance behaviors. To test the proposed relationship, we set up the data with "state × week" as the unit of analysis, by aggregating the computed data at the state level (50 states and the District of Columbia) and split the time series data into eight weeks (from Week 1: March 1-7, 2020, to Week 8: April 18-25, 2020). The values of political leaning of different states were held constant as they would not change over the analysis time period. This procedure generated 408 units, each of which was comprised of a set of computed values representing state-level public's emotions, information-seeking, information-sharing, and compliance behaviors, and political leaning.

Our first hypothesis (H1) is that the more low-certainty emotions people experienced during the COVID-19 outbreak, the more likely they would seek COVID-related information; and the more high-certainty emotions people experienced, the less likely they would seek information. To test this hypothesis, the percentages of fear and sadness were combined to form the extent of low-certainty emotion, and the percentage of anger was used to indicate the degree of high-certainty emotion. Then, a regression analysis was conducted with the aggregated data at the state and the weekly levels. The percentages of high and low-certainty emotions were included as the two continuous independent variables, the standardized information-seeking behavior score was entered as the dependent variable, and political leaning as the control variable. To summarize:

**H1**: During the COVID-19 outbreak, the public's low-certainty emotions are positively related to their information seeking behaviors, while the public's high-certainty emotions are negatively related to their information-seeking behaviors.

*Regression Equation for H1 testing:*
{information-seeking behavior} = α + β1{high-certainty emotion} + β2{low-certainty emotion} + β3{political leaning} + ϵ

Based on the Appraisal Tendency Framework, our second hypothesis (H2) is that the more agonistic emotion (e.g., anger) people felt during the COVID19 outbreak, the more likely they would transmit COVID-related information; and the more passive emotion (e.g., fear or sadness) people felt, the less likely they would transmit information. To test H2, a regression analysis was conducted with the standardized information-transmitting score as the dependent variable, the percentages of high and low-certainty emotions as independent variables, and political leaning as the control variable. To summarize:

**H2**: During the COVID-19 outbreak, the public's agnostic emotions are positively related to their information-sharing behaviors, while the public's passive emotions are negatively related to their information-sharing behaviors.

*Regression Equation for H2 testing:*
{information-sharing behavior} = α + β1{agnostic emotion} + β2{passive emotion} + β3{political leaning} + ϵ

Our third hypothesis (H3) is that the more low-certainty emotion (e.g., fear or sadness) people experienced during the COVID-19 outbreak, the more likely they would comply with the social distancing order. In contrast, the more high certainty emotion (e.g., anger) people experienced, the less likely they would comply. To test this hypothesis, two rounds of regression analyses were performed, with the standardized score of movements in residential places and the standardized score of movements in public places as the dependent variable in each of the regression models, and the same independent and control variables as before. Hence,

**H3**: During the COVID-19 outbreak, the public's low certainty emotions are positively related to their compliance with stay-at-home orders, while the public's high-certainty emotions are negatively related to their compliance with stay-at-home orders.

*Regression Equation for H3 testing:*
{Residential places mobility} = α + β1{high-certainty emotion} + β2{low-certainty emotion} + β3{political leaning} + ϵ

{Public places mobility} = α + β1{high-certainty emotion} + β2{low-certainty emotion} + β3{political leaning} + ϵ

Figure 1 shows the variables considered for this study and figure 2 shows the variables considered for each hypothesis.

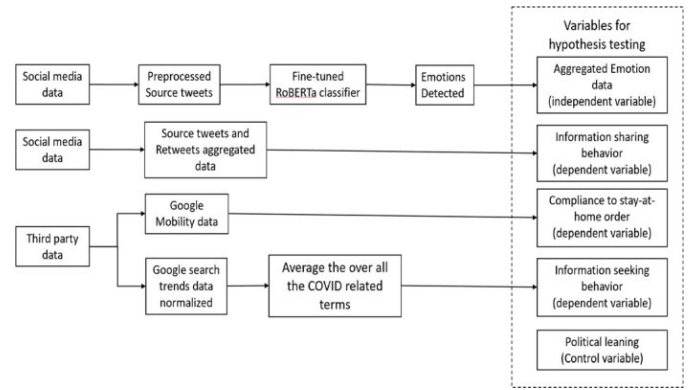

Figure 1: Variables for the Study

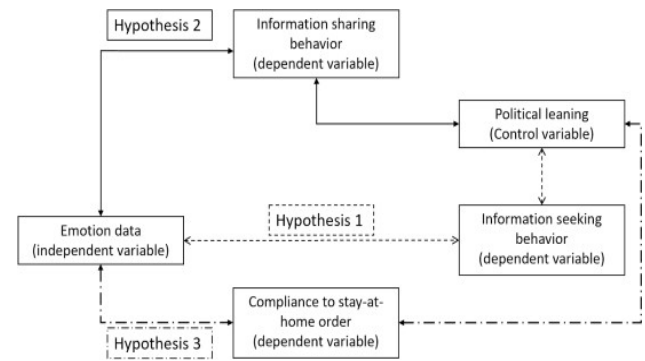

Figure 2: Figural Representation of the Study Hypotheses.

## Results

We first examine the primary emotion experienced by the people in source tweets and variations of these emotions during the outbreak of COVID-19. Figure 3 show the percentages of emotions detected in the social media data over the 8-weeks. The results show that anger was the most dominant emotion experienced (3.39%) which is closely followed by fear (3.30%) and positive emotion (2.94%) by the public during the outbreak of COVID-19. The least dominant emotion was sadness (1.84%) during the early weeks of the COVID19 breakout. Figure 3 also illustrates the changes in percentages of each discrete emotion over the eight-week time period. We can observe that anger fluctuated considerably during the breakout: It reached a relatively high point in Week 3 and then decreased to its lowest point in Week 5; however, the relative prevalence of anger jumped to new heights in Weeks 6 and 7. Fear increased from Week 1 to Week 6 with some fluctuations and became the most prevalent emotion in Week 6. Sadness increased gradually from Week 1 to Week 6, before slightly decreasing in Week 7 and Week 8, and it remained the least preva-lent emotion over the eight weeks. Finally, for positive emotion, it gradually increased from Week 1 to Week 3 and remained steady in the following five weeks. Interestingly, from Week 3 to Week 5, positive emotion was as prevalent as each discrete negative emotion.

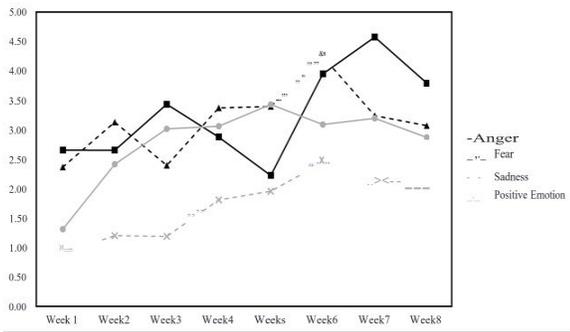

Figure 3: Discrete Emotions Detected in the Tweets

In general, the data reveal two patterns of the public's emotions. First, as the COVID-19 pandemic developed and multiple crisis events occurred along the way, people were experiencing a combination of negative emotions with varying degrees. Second, in some weeks, the high-certainty and agonistic emotion (anger) was relatively more prevalent whereas in some other weeks the low-certainty and passive emotions (fear and sadness) were more prevalent. This is probably due to how specific crisis events and response measures were appraised by the public. It is also noteworthy that, while positive emotion was not common at the beginning of the crisis, people had an increasing tendency to express positive emotion as the crisis evolved.

The results for our first hypothesis (H1) show that there is a significant impact of low-certainty emotions on the public's information-seeking behavior. Table 1 shows that the low-certainty ($\beta = .12$, $t = 2.37$, $p = .018$) emotion is a significant positive predictor of level of information seeking ($F[3, 404] = 4.95$, $p = .002$, $R^2 = .03$) while high-certainty emotion ($\beta = -.05$, $t = -.97$, $p = .331$) is not a significant predictor of information seeking.

| Variables | B | SE | β | t | p |
|---|---|---|---|---|---|
| High-certainty emotion | -36.05 | 37.07 | -.05 | -.97 | .331 |
| Low-certainty emotion | 70.05 | 29.52 | .12 | 2.37 | .018 |
| Political leaning | 18.29 | 6.24 | .14 | 2.93 | .004 |

Table 1: Regression Analysis Summary for Emotions Predicting Information-Seeking Behaviors

The results for second hypothesis (H2) show that when political leaning is controlled, the passive emotions ($\beta = .38$, $t = 8.15$, $p < .001$) expressed by the public was a significant predictor of the level of information sharing ($F[3, 404] = 23.65$, $p < .001$, $R^2 = .14$). However, the extent of agonistic emotion was not a significant predictor of information sharing ($\beta = .03$, $t = .73$, $p = .464$) (see Table 2). Hence, the result shows a pattern contrasting our predicted H2, since the more passive emotion the public experienced during the COVID-19 outbreak, the more information sharing they engaged in.

| Variables | B | SE | β | t | p |
|---|---|---|---|---|---|
| Agonistic emotion | 1.68 | 2.29 | .03 | .73 | .464 |
| Passive emotion | 14.85 | 1.82 | .38 | 8.15 | <.001 |
| Political leaning | .05 | .39 | .01 | .12 | .905 |

Table 2: Regression Analysis Summary for Emotions Predicting Information-Transmitting Behaviors

The results for our final hypothesis (H3) show that, low-certainty ($\beta = .33$, $t = 7.24$, $p < .001$) and high-certainty emotions ($\beta = .12$, $t = 2.54$, $p = .011$) are both significant positive predictors of the public's movement in residential areas, when political leaning is controlled ($F[3, 404] = 30.42$, $p < .001$, $R^2 = .18$) (see Table 3). Meanwhile, the results also show that, both high certainty ($\beta = -.11$, $t = -2.45$, $p = .015$) and low certainty emotions ($\beta = -.34$, $t = -7.51$, $p < .001$) are negative predictors of the public's movement of in public areas when political leaning is controlled ($F[3, 404] = 36.91$, $p < .001$, $R^2 = .21$) (see Table 4). To summarize H3, we can say that both low-certainty and high-certainty emotions are related to the public's compliance to stay-at-home order. The relatively larger coefficients of the lower-certainty emotion variable indicate that the extent of low-certainty emotion might be a more prominent positive predictor of the public's compliance behavior, which is in line with our general prediction. Table 3 and 4 show the regression Analysis summary for emotions predicting movements in residential areas and public areas respectively.

| Variables | B | SE | β | t | p |
|---|---|---|---|---|---|
| High-certainty emotion | 46.22 | 18.19 | .12 | 2.54 | .011 |
| Low-certainty emotion | 104.86 | 14.48 | .33 | 7.24 | <.001 |
| Political leaning | 15.75 | 3.06 | .23 | 5.15 | <.001 |

Table 3: Regression Analysis Summary for Emotions Predicting Movements in Residential Areas

| Variables | B | SE | β | t | p |
|---|---|---|---|---|---|
| High-certainty emotion | -104.80 | 42.70 | -.11 | -2.45 | .015 |
| Low-certainty emotion | -255.24 | 34.01 | -.34 | -7.51 | <.001 |
| Political leaning | -47.07 | 7.19 | -.29 | -.6.55 | <.001 |

Table 4: Regression Analysis Summary for Emotions Predicting Movements in Public Areas

## Application

Despite the topic's growing importance, research on the role and impact of the public's emotions on their coping behaviors in public health crises has been limited, and analysis of behaviors in a natural setting is notably lacking. By applying a computational model (i.e., fine-tuned RoBERTa) and connecting it with third-party behavioral data, we demonstrate the feasibility and reliability of our proposed approach in providing a large-scale examination of the public's emotions and coping behaviors in a real-world crisis. This computational approach complements prior crisis communication research based on self-reported, small-scale experiments and survey data. The survey data being self-reported and experimental data collected from a controlled environment are not scalable and reliable. Hence, our approach is beneficial during longitudinal crises where the data about the public's emotions and behavior is hard to understand with surveys and experimental designs.

This approach can help practitioners better understand how to segment the affected population in public health crises based on their emotions and tailor crisis response strategies to address the prevalent emotions in each segment. Assessing the public's emotion fluctuations and the connections with their coping behaviors using real world data can greatly improve the effectiveness of crisis responses.

Interestingly, in this study, while the results from largescale, naturalistic data provides valuable insight into a real world public health crisis, they are not entirely consistent with findings from previous surveys and experimental studies on relevant topics (e.g., Jin, Fraustino and Liu 2016). Strategic communication in disease outbreak crises is a growing multidisciplinary research field that intersects crisis communication, risk communication, and health communication. These inconsistent findings point to a promising future research direction to bridge the different theoretical and methodological approaches in these different fields.

## Limitation

The main limitation of this study is that we have not cross validated this study's findings using highly controlled experimental methods. Such cross-validation research can help further improve the accuracy of the computational analysis of psychological variables. Secondly, we have considered a limited number of emotions. Studies have shown that the public goes through different emotions like anxiety, despair, hope, and sympathy during a crisis. We will involve datasets that include other positive and negative emotions during a crisis.

Additionally, this study focused on the first eight weeks of the COVID-19 outbreak, and thus, the research findings are limited within this time frame. In our future work, we plan to develop a more in-depth model to test further how the relationships between different emotions and coping behaviors may change over time as a longitudinal crisis evolves into chronic and resolution phases.

Currently, this approach uses publicly available data and reports. In the future, if this approach is implemented in real world practices, related ethical concerns also must be addressed. Informed consent must be provided to the involved users regarding how the data may be used.

## Conclusion

In this study, we analyzed the public's most prevalent emotions during the COVID-19 outbreak and the impacts on their coping behaviors. We found that anger and fear were prominent emotions experienced by the public during the studied period. Also, the extent of low-certainty and passive discrete emotions was related to increased information seeking and information-sharing behaviors. In addition, both high and low-certainty emotions were positively related to the public's compliance with stay-at-home orders. This study demonstrates the feasibility of using a fine-tuned RoBERTa model to detect the public's emotions in public health crises and using third-party data to assess the impacts of emotions on coping behaviors. This approach, when practically implemented, can help crisis management professionals understand the public's real-time emotion fluctuations and inform their strategies to facilitate the public's appropriate coping behaviors.